\documentstyle[emulateapj,psfig,apjfonts]{article}
\def\gs{\mathrel{\raise0.35ex\hbox{$\scriptstyle >$}\kern-0.6em 
\lower0.40ex\hbox{{$\scriptstyle \sim$}}}}
\def\ls{\mathrel{\raise0.35ex\hbox{$\scriptstyle <$}\kern-0.6em 
\lower0.40ex\hbox{{$\scriptstyle \sim$}}}}
\submitted{Received: 2000 March 26; accepted 2000 May 9}
\lefthead{Ivison et al.}
\righthead{A candidate submm proto-cluster at $z$ = 3.8}

\begin{document}

\title{An excess of submm sources near 4C\,41.17: a
       candidate proto-cluster at $z$=3.8?}

\author{R.\,J.\ Ivison,$\!$\altaffilmark{1,2}
        J.\,S.\ Dunlop,$\!$\altaffilmark{3} Ian Smail,$\!$\altaffilmark{4,5}
        Arjun Dey\altaffilmark{6},
        Michael C. Liu\altaffilmark{7} \& J.\,R.\ Graham\altaffilmark{7}
        }
\affil{\small 1) Department of Physics \& Astronomy, University College
                 London, Gower Street, London WC1E 6BT, UK}
\affil{\small 3) Institute for Astronomy, University of Edinburgh, Blackford
                 Hill, Edinburgh EH9 3HJ, UK}
\affil{\small 4) Department of Physics, University of Durham, South Road, 
                 Durham DH1 3LE, UK}
\affil{\small 6) NOAO, 950 N.\ Cherry Ave, Tucson AZ 85719}
\affil{\small 7) Astronomy Department, University of California at Berkeley,
                 CA 94720}

\setcounter{footnote}{5}

\altaffiltext{2}{PPARC Advanced Fellow.}
\altaffiltext{5}{Royal Society University Research Fellow.}

\begin{abstract}
Biased galaxy-formation theories predict that massive galaxies at high
redshifts should act as signposts to high-density environments in the
early universe, which subsequently evolve into the cores of the
richest clusters seen at the present day. These regions are
characterised by over-densities of young galaxies, perhaps including a
population of dusty, interaction-driven starbursts --- the progenitors
of massive cluster ellipticals. By searching for this population at
submillimeter (submm) wavelengths we can therefore test both galaxy-
and structure-formation models. We have undertaken such a search in
the field of a $z=3.8$ radio galaxy, 4C\,41.17, with the SCUBA submm
camera. Our extremely deep 450- and 850-$\mu$m maps reveal an
order-of-magnitude over-density of luminous submm galaxies compared to
typical fields (the likelihood of finding such an over-density in a
random field is $<2 \times 10^{-3}$). The SCUBA galaxies have
bolometric luminosities, $>10^{13}\,$L$_\odot$, which imply
star-formation rates (SFRs) consistent with those required to form a
massive galaxy in only a few $10^8$ years.  We also note that this
field exhibits an over-density of extremely red objects (EROs), some
of which may be associated with the submm sources, and Lyman-break
galaxies. We propose that the over-densities of both submm and ERO
sources in this field represent young dusty, starburst galaxies
forming within a proto-cluster centered on the radio galaxy at
$z=3.8$, which is also traced by a less-obscured population of
Lyman-break galaxies.
\end{abstract}



%
%
%
\section{Introduction}

Galaxy-formation theories are developing rapidly and have claimed some
success at reproducing the properties of galaxies in the local
universe (Cole et al.\ 1994). However, at the moment their predictions
of the early evolution of galaxies are comparatively
untested. Moreover, the areas where the theoretical predictions are
most reliable tend to be those where the observational tests are most
difficult at high redshift, e.g., the evolution in the mass function
of galaxies (White \& Frenk 1991).  One testable prediction of
hierarchical galaxy-formation models at high redshifts deals with the
clustering behaviour of massive galaxies at early epochs: such
galaxies are expected to cluster strongly in regions of highest
density, areas which should subsequently form the cores of massive
clusters of galaxies in the local universe (Kaiser 1984; Baron \&
White 1987; Efstathiou \& Rees 1988; Kauffmann et al.\ 1999). Thus by
searching for over-densities of massive galaxies in the distant
universe we can test the predictions of hierarchical models {\it as
well as} investigating the formation and evolution of the luminous
galaxies seen in present-day clusters.  However, to achieve this we
must find a method of selecting the densest regions in the early
universe.

One possible technique is to use massive galaxies as signposts for
high-density environments at high redshifts.  Observations of distant,
luminous radio galaxies have led to the suggestion that they are
massive ellipticals (Matthews, Morgan \& Schmidt 1964; Lilly \&
Longair 1984), a belief confirmed recently by {\it Hubble Space
Telescope} ({\it HST}) studies which show they possess $r^{1/4}$-law
profiles characteristic of elliptical galaxies (McLure et al.\ 1999;
Zirm et al.\ 2000).  Locally these massive elliptical galaxies typical
reside in galaxy clusters and hence luminous radio galaxies should
make good markers to search for the progenitors of rich galaxy
clusters at high redshifts.

The picture of hierarchical formation of massive galaxies and their
environments is supported on small scales by recent rest-frame optical
imaging of radio galaxies (van Breugel et al.\ 1998; Pentericci et
al.\ 1998): at the earliest epochs ($z\ge 4$) there is evidence of
diffuse emission on large scales (5--10$\,$$''$) and sub-clumps
similar in scale ($\sim10$\,kpc) to radio-quiet star-forming galaxies;
these then appear to evolve into more compact structures by $z\sim
2$. Interestingly, on somewhat larger scales there is also an apparent
10--100-fold increase in the surface density of EROs --- some of which
are dusty, star-forming galaxies akin to local ULIRGs (Dey et al.\
1999; Smail et al.\ 1999) --- around high-redshift radio galaxies
(HzRGs) and quasars as compared with the field (Hu \& Ridgway 1994;
Elston, Rieke \& Rieke 1988; Arag\'on-Salamanca et al.\ 1994; Dey,
Spinrad \& Dickinson 1995; Yamada et al.\ 1997).  On similar scales,
the distorted radio morphologies and high rotation measures of some
HzRGs suggest they reside in high-density environments (Carilli et
al.\ 1997) with extended X-ray emission detected around one example,
1138$-$262 at $z=2.2$ (Carilli et al.\ 1998).

%
%
\begin{figure*}
\centerline{\psfig{file=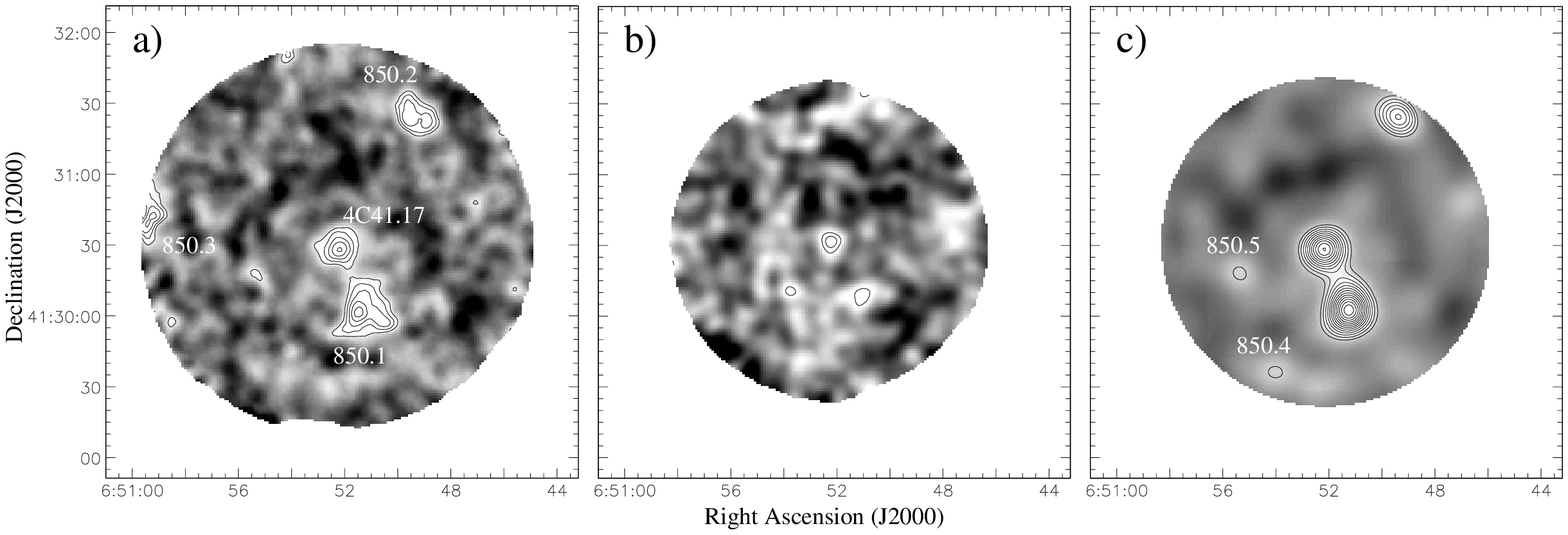,width=7.3in,angle=0}}

\noindent{\scriptsize
{\sc Fig.~1.}---a) Full 850-$\mu$m image of the $z=3.8$ radio galaxy,
4C\,41.17, at the original $\sim$14$''$ resolution. Negative features
are due to the chopped/nodded observing procedure. Contours are plotted
at $3, 4, 5, 6, 7 \times 1.5\,$mJy\,beam$^{-1}$. b) 450-$\mu$m image of
4C\,41.17, smoothed to a resolution of 10$''$ FWHM.  Contours are
plotted at $3, 4 \times 8.6\,$mJy per 10$''$ beam$^{-1}$. c) Central,
cleaned portion of the 850-$\mu$m image.  Contours are plotted at $3, 4
\ldots 15 \times 0.8\,$mJy\,beam$^{-1}$, where the beam has been smoothed
to 20$''$ in the background regions.

\vspace*{-1mm}
}
\end{figure*}

Confirming that HzRGs reside in high-density environments would also
provide substantial insight into galaxy evolution in these regions,
especially the formation of giant ellipticals in the hierarchical
models already discussed. Optical/IR studies have targeted the
evolution of ellipticals in clusters at $z < 1$ (Arag\'on-Salamanca et
al.\ 1993; Stanford, Eisenhardt \& Dickinson 1998; Bower, Lucey \&
Ellis 1992; Ellis et al.\ 1997); theoretical work has investigated
their evolution in the context of hierarchical formation in
high-density regions (Kauffmann \& Charlot 1998; Baugh et al.\
1999). Pinpointing the progenitors of ellipticals at very high
redshifts would allow straightforward tests of the latter models and
would enable the time-line of the optical/IR cluster studies to be
extended beyond $z\sim 1.3$ (Stanford et al.\ 1997, 1998; Rosati et
al.\ 1999; Liu et al.\ 2000).

In the hierarchical picture, lower-mass structures will accrete onto
the high-mass peak of the proto-cluster identified with the radio
galaxy, the final assembly of this structure achieved through a series
of mergers which will strongly affect the galaxies residing within the
dark matter halo, triggering intense starbursts analogous to mergers
in the local universe (Sanders \& Mirabel 1996). The dust created in
these starbursts will absorb a considerable fraction of the UV/optical
light emitted by young stars, re-emitting it in the rest-frame
far-IR. An extreme and highly obscured starburst with a SFR of $\sim
1000\,$M$_\odot$yr$^{-1}$ would have a bolometric luminosity, $L_{\rm
bol}$, of $\sim 10^{13}\,$L$_\odot$, and the majority of this would
appear at rest-frame far-IR wavelengths (Ivison et al.\ 1998).

The negative $K$-correction for dust emission in the submm passband
(Blain \& Longair 1993) means that a luminous, dusty starburst with
$L_{\rm bol} \sim 10^{13}\,$L$_\odot$ would have an 850-$\mu$m flux of
$\sim 10\,$mJy if observed at {\it any} redshift between 1 and 10. The
advent of sensitive arrays working at submm wavelengths, in particular
the SCUBA camera (Holland et al.\ 1999) on the James Clerk Maxwell
Telescope (JCMT) have enabled efficient and sensitive surveys of dusty
starbursts,  probing out to  high redshifts (Smail, Ivison \& Blain
1997; Barger et al.\ 1998; Hughes et al.\ 1998; Eales et al.\ 1999).

Thus by using SCUBA to undertake targeted observations of the
environments of known HzRGs, searching for over-densities of bright
submm sources, we can test whether HzRGs are located in the cores of
rich proto-clusters {\it and} constrain the formation epoch of massive
cluster galaxies.  For targets at $z\gs 2$ the resolution and field of
view of SCUBA at 850$\,\mu$m means we are sensitive to star-forming
galaxies distributed on scales from 100 to 1000\,kpc --- well-matched
to the predicted virial radii of the most massive clusters at these
epochs (Jenkins et al.\ 2000).

In this paper we present deep 450- and 850-$\mu$m maps of one of the
most distant and powerful known radio galaxies, 4C\,41.17 at $z=3.8$
(for which 1$''$ = 6.6\,$h_{50}^{-1}$\,kpc, with $h_{50} =
H_0/50$\,km\,s$^{-1}$\,Mpc$^{-1}$ and $q_0=0.5$).  There is already
some evidence for a massive structure associated with 4C\,41.17.
Based upon a Lyman-break search using $UVR$ imaging, Lacy \& Rawlings
(1996) have suggested that there is a modest excess of high-redshift
galaxies in this region, having found six candidate $z\gs 3.4$
galaxies in a 1.5-arcmin$^2$ field, roughly centered on 4C\,41.17.
They show that this density is slightly higher than that expected from
the field density found by Steidel, Pettini \& Hamilton (1995), but
the complications introduced by comparing their $UVR$ study with
Steidel et al.'s $\mathcal{UGR}$ survey, as well as issues of cosmic
variance (Steidel et al.\ 1999), make it difficult to assess the
over-density.  Other evidence for a deep potential centered on the
radio galaxy includes the presence of an extended Ly$\alpha$ halo
reaching to $>$100\,kpc (Chambers, Miley \& van Breugel 1990; Dey
1999).

In the next section we present our submm, near-IR and optical
observations of this field and their reduction. We discuss our analysis
and results, in the context of supporting archival data at other
wavelengths, in \S3 and give our conclusions in \S4.

\section{Observations and Data Reduction}

\subsection{Submillimeter Observations}

During 1998 October we used SCUBA to map a $\sim 2.5'$-diameter field
at 450- and 850-$\mu$m, centred on 4C\,41.17 (Fig.~1). The secondary
mirror followed a jiggle pattern designed to fully sample the image
plane, chopping by 45$''$ between the `on' and `off' positions at
7\,Hz whilst the telescope nodded between the same positions every
16\,s in an on--off--off--on pattern. The chop throw was chosen to
optimise calibration accuracy, minimise noise, maximise regions for
which both `on' and `off' beam information would be available, and
minimise the possibility of bright sources close to the field of view
contaminating the map. Two directions were used for the chopping:
north--south (N--S) as the target rose and set; east-west (E--W) when
the target was near transit. This method ensured the chop direction
was close to azimuthal whilst retaining the advantages of on-array
chopping. It also provided independent maps which could be cleaned
separately and coadded, reducing the risk of spurious detections and
of chopping consistently from one source onto another.

%
%
\begin{figure*}
\centerline{\psfig{file=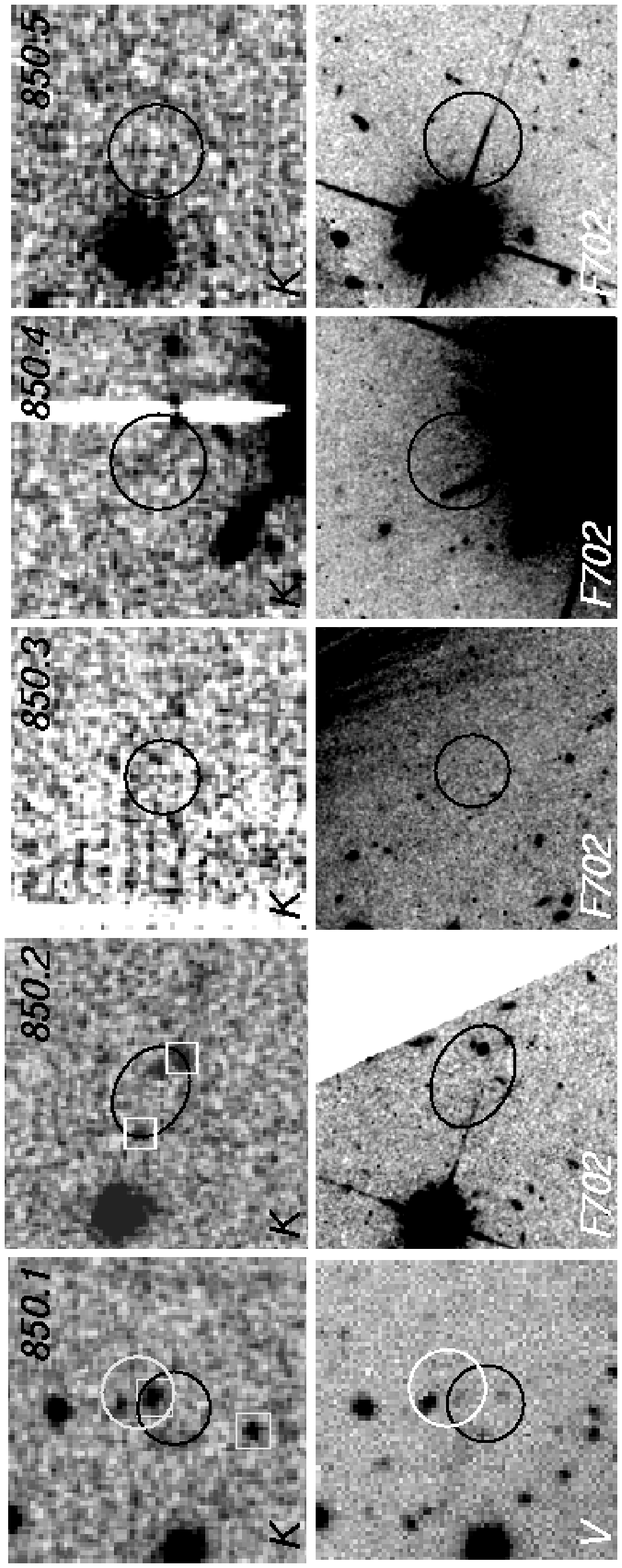,width=7.3in,angle=270}}

\noindent{\scriptsize {\sc Fig.~2.}--- KPNO 2.1-m $K'$-band imaging
and {\it HST} F702W images of $25'' \times 25''$ regions around the
submm sources, smoothed with 0.5$''$- and 0.15$''$-FWHM Gaussians at
$K'$ and $F702W$ respectively. For HzRG850.1, the optical data are
from Keck\,{\sc ii} (\S2.2). EROs are marked with squares; 850-$\mu$m
positions are marked with 6$''$- or 8$''$-diameter black circles (an
ellipse for HzRG850.2); the 450-$\mu$m position of HzRG850.1 is marked
with a white circle.

\vspace*{-1mm}
}
\end{figure*}

The sky was exceptionally stable and transparent for the five nights
during which data were obtained. The total exposure time was 36\,ks
and the opacity at 850\,$\mu$m, measured every hour, was typically
0.15--0.20, never rising above 0.25. Flux calibration was accomplished
using beam maps of Mars. Nightly calibration factors were stable:
consistent at the 5\% level (r.m.s.)  at 850\,$\mu$m.  Even at
450\,$\mu$m, factors were stable at the 10\% level. Absolute flux
calibration should be accurate to $\sim 10$\%.

Maps were created using {\sc surf} (Jenness \& Lightfoot 1997). In
order to extract reliable source positions and flux densities, the
images (Fig.~1) and the symmetric $-1, +2, -1$ zero-flux beam (which
arises from chopping and nodding) were deconvolved using a modified
version of the {\sc clean} algorithm (Hogbom 1974) as follows.  The
E--W- and N--S-chopped maps were treated separately for the purpose of
deconvolution.  First, both the image and the appropriate beam (from
contemporaneous maps of blazars) were convolved with a Gaussian (FWHM
14$''$, similar to the core of the JCMT beam at 850\,$\mu$m). The
smoothed image was then cleaned using the smoothed beam, down to a
level equivalent to twice the r.m.s.\ noise of the smoothed data,
using a loop-gain of 0.1. This has the effect of gradually removing
the positive core of a real source from the data, while at the same
time filling in its negative sidelobes. When the process is complete,
the original image is reduced to a map of background noise (still
containing real sources which are too faint to have been identified as
significant peaks by the {\sc clean} algorithm) while the absolute
values of removed flux are stored as a set of delta functions (in the
case of unresolved sources) located at the positions of the original
source peaks.

The restored images were produced by halving the size of the delta
functions then convolving them with a Gaussian (FWHM 14$''$) and
adding them back into the residual noise image left by the cleaning
process. The result is a restored image with the angular resolution of
the original but free from negative sidelobes associated with
significant sources. One advantage of this process is that it should
resurrect any significant sources whose flux has been depressed (or
annihilated) because they happen to lie in the negative sidelobe of
another source, though only sources within 70$''$ of the field centre
are treated properly by the cleaning procedure. Sources which were
produced in both the E--W- and N--S-chopped, restored sub-images, with
positions differing by less than 3$''$, can be regarded as robust
detections.

\subsection{Near-Infrared and Optical Observations}

We used the Ohio State/NOAO IR Spectrometer (ONIS) on the 2.1-m
telescope of the Kitt Peak National Observatory to obtain $K$-band
images of the 4C\,41.17 field covering a $5.6'\times8.0'$ field at
0.34\arcsec/pixel sampling. The observations were taken in 0.9\arcsec\
seeing on the nights of 1998 November 14, 15 and 1999 March 20.
Observations were made using a random dither pattern, with multiple
60\,s integrations at each pointing and a combined exposure time of
16.3\,ks.  The data were dark subtracted, linearized, flat fielded,
and the final mosaic (Fig.~2) was constructed using a modified version
of {\sc dimsum} in {\sc iraf}\footnote{{\sc iraf} is distributed by
the National Optical Astronomy Observatories.}. The photometric
calibration was accomplished using observations of several {\it HST}
IR standard stars (Persson et al.\ 1998) and checked against the
photometry of Graham et al.\ (1994) for the sources in common. The
images were astrometrically calibrated using the U.S.\ Naval
Observatory A2.0 Catalog.  The final $K$-band mosaic has a seeing FWHM
of 1.0\arcsec\ and reaches a $5\sigma$ limit of $K=19.2$ in a
2\arcsec\ diameter aperture.

In addition to these wide-field observations, we also obtained deep
$JHK$ images of the fields of two of the brighter SCUBA sources
(HzRG850.1 and HzRG850.2; see below) using the Near-IR Camera (NIRC,
Matthews \& Soifer 1994) at the W.\,M.\ Keck Observatory in
photometric conditions on the nights of 1999 March 24 and 25.  NIRC
has a 38\arcsec\ field of view with 0.15\arcsec/pixel sampling and the
observations were taken with a random dither pattern, with a 120-s
exposure per pointing.  Over the two nights, we obtained total
integrations on HzRG850.1 of 1.2\,ks, 2.4\,ks and 1.2\,ks in $JHK$
respectively, with 2.4\,ks, 1.2\,ks and 2.4\,ks in $JHK$ on the
HzRG850.2 field.  The frames were reduced and combined in a standard
manner and we show the resulting composite color images in Fig.~3.
The seeing measured off the final stacked frames is 0.5--0.6\arcsec
and the frames were calibrated using the faint standards from Persson
et al.\ (1998) yielding detection limits of $J\sim 24$, $H\sim 22.5$
and $K\sim 21$.

$V$-band images of the 4C\,41.17 field were obtained using the Low
Resolution Imaging Spectrometer (LRIS; Oke et al. 1995) on Keck\,{\sc
ii} on 1998 September 19.  The total exposure time was 2.7\,ks, broken
into 9 dithered sub-exposures.  Conditions were non-photometric, but
the data were calibrated using magnitudes of objects in the field from
Lacy \& Rawlings (1996). The final image reaches a $5\sigma$ limiting
magnitude of $V=26.3$ in a 2\arcsec\ diameter aperture.

Finally, we have also made use of the archival {\it HST} WFPC2 image
of this field (GO\#5511).  This comprises a total of 21.6\,ks
integration through the F702W filter and reaches a 3-$\sigma$ limiting
magnitude of $R=26.0$ (2\arcsec\ aperture, converted to Cousins $R$
using photometry of sources with $R-K\sim 4$ in Graham et al.\ 1994).

These combined optical and near-IR observations identify several new
ERO sources to add to the two originally found by Graham et al.\
(1994).  In total we identify at least five objects with extreme
colors, $R-K\geq 6$ or $V-K\geq 7$, brighter than $K=20$ within
roughly a 2 arcmin$^2$ field.  This surface density is nearly an order
of magnitude higher than that seen in blank fields (e.g.\ Thompson et
al.\ 1999) and as we discuss below several of these objects maybe
associated with the submm sources in this field.

%
%
\begin{figure*}

\centerline{\psfig{file=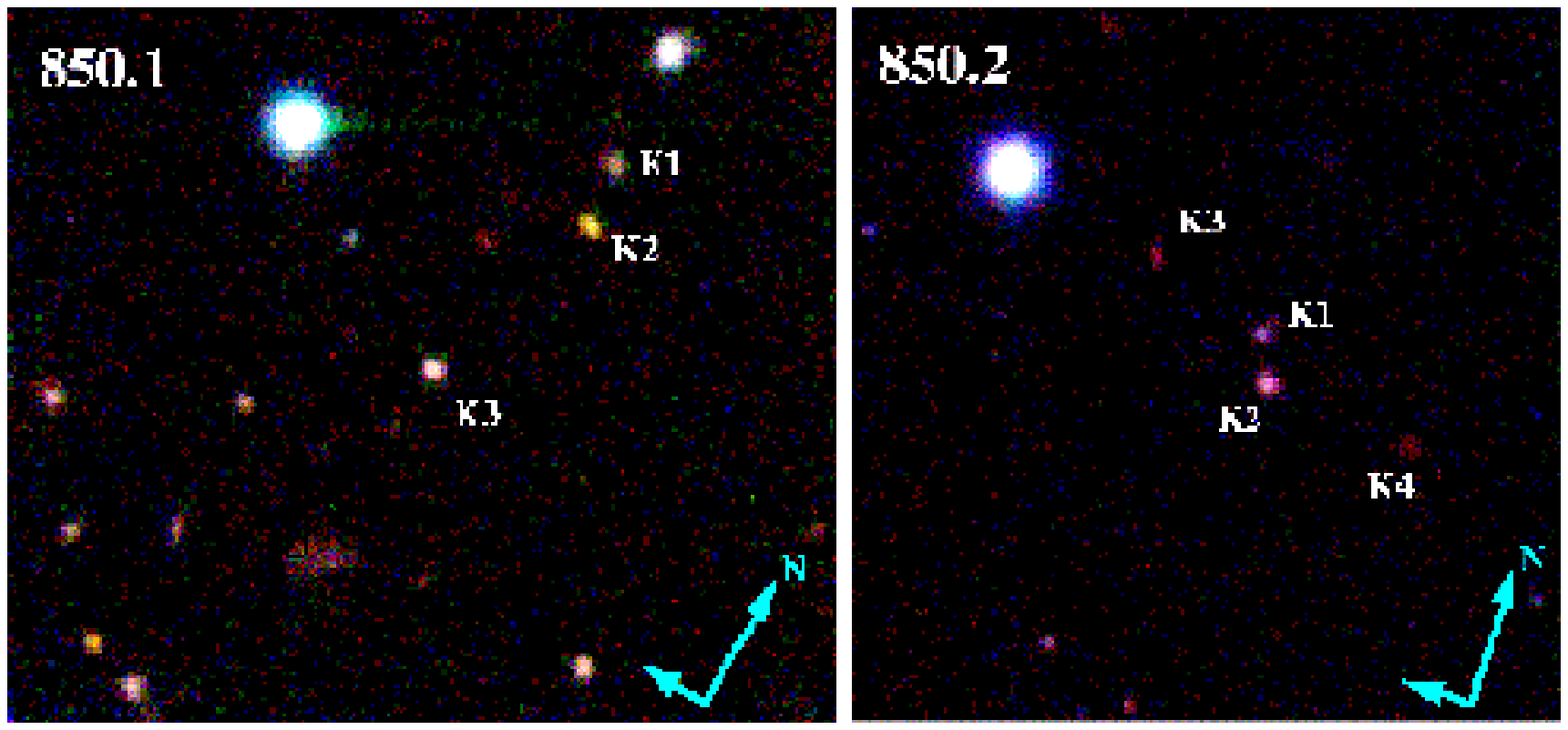,width=7.4in,angle=0}}

\noindent{\scriptsize {\sc Fig.~3.}--- Composite color images of the
fields of HzRG850.1 and HzRG850.2, constructed from the Keck $JHK$
imaging.  We label the possible candidate counterparts to the submm
sources as well as several other sources in the close vicinity.  The
field of HzRG850.1 is $54\arcsec \times 47\arcsec$, while HzRG850.2 is
47\arcsec\ square. Orientations are shown.

\vspace*{-1mm}
}
\end{figure*}

\section{Results and Discussion}

\subsection{Submillimeter sources}

At 850\,$\mu$m, the cleaned portion of the map (4.3\,arcmin$^2$)
reaches a 4-$\sigma$ sensitivity of 3.2\,mJy\,beam$^{-1}$
(6.0\,mJy\,beam$^{-1}$ for a further 1.6\,arcmin$^2$), sufficient to
detect galaxies with SFRs of a few 100\,M$_\odot$ yr$^{-1}$ at the
redshift of 4C\,41.17.

In the cleaned image we detect three sources well above the 4-$\sigma$
flux limit with positions differing by $<3''$ in the sub-images
(Table~1); of these, one is associated with the radio galaxy, as
determined previously by Dunlop et al.\ (1994). The others are new
detections and one of them, HzRG850.1, is the brightest 850-$\mu$m
source yet reported. As we discuss in \S3.2, this represents a large
over-density compared to the expected blank-field count of 0.1--0.3
per SCUBA field.

%
%
{\scriptsize
\begin{center}
{\sc Table 1}

\vspace{0.1cm}
{\sc Positions and Flux Densities of Submm Sources}

\vspace{0.3cm}
\begin{tabular}{lllrc}
\hline\hline
\noalign{\smallskip}
{IAU} &{~~~~~~~~RA$^{\rm a}$} &{~~~~~~Dec.$^{\rm a}$} &{$S_{850\mu{\rm m}}^{\rm b}$}~~ &{$S_{450\mu{\rm m}}^{\rm b}$}~~  \cr
{name}&{~~~(h min s $\pm$ s)} &{~~($^{\circ}$ $'$ $''$ $\pm$ $''$)}&{(mJy)}~ & {(mJy)}~ \cr
\hline
\noalign{\smallskip}
HzRG850.1$^{\rm c}$         &
$06\ 50\ 51.2\ \pm 0.3$     &
$+41\ 30\ 05\ \pm 3$        &
$15.6 \pm 1.8$              &
$34.1 \pm 9.3$              \cr
                            &
$06\ 50\ 51.0\ \pm 0.3^{\rm d}$ &
$+41\ 30\ 08\ \pm 3^{\rm d}$&
                            &
                            \cr
HzRG850.2$^{\rm c}$         &
$06\ 50\ 49.3\ \pm 0.4$     &
$+41\ 31\ 27\ \pm 3$        &
$8.7 \pm 1.2$               &
---$^{\rm e}$               \cr
HzRG850.3$^{\rm f}$         &
$06\ 50\ 59.3\ \pm 0.4$     &
$+41\ 30\ 45\ \pm 3$        &
$6.5 \pm 1.6$               &
---$^{\rm e}$               \cr
4C\,41.17$^{\rm c}$         &
$06\ 50\ 52.14$             &
$+41\ 30\ 30.8$             &
$11.0 \pm 1.4$              &
$35.3 \pm 9.3$              \cr
HzRG850.4                   &
$06\ 50\ 54.0\ \pm 0.4$     &
$+41\ 29\ 39\ \pm 4$        &
$2.8 \pm 0.8$               &
$3\sigma < 26$              \cr
HzRG850.5                   &
$06\ 50\ 55.3\ \pm 0.4$     &
$+41\ 30\ 20\ \pm 4$        &
$2.4 \pm 0.8$               &
$3\sigma < 26$              \cr
\noalign{\smallskip}
\noalign{\hrule}
\noalign{\smallskip}
\end{tabular}
\end{center}

\vspace*{-2mm}
\noindent
{\sc Notes:} $a$) 850-$\mu$m position (J2000); $b$) Photometry measurements,
taking the 15$''$-diameter aperture-corrected value at 850\,$\mu$m and
the 7.5$''$-diameter aperture-corrected value at 450\,$\mu$m; errors
include a 10\% contribution from the uncertainty in absolute flux
calibration; $c$) Positions and fluxes were measured from the cleaned
image. The 850-$\mu$m maps were shifted 0.3$''$ west, 2.4$''$ north
(1.0$''$ west, 0.4$''$ south for the 450-$\mu$m map) such that the
submm emission from 4C\,41.17 coincided with the radio core; $d$)
450-$\mu$m position; $e$) Source lies off the 450-$\mu$m map; $f$)
Lies outside of 70$''$-radius cleaned zone.

}
\vspace*{2mm}

The flux measurements of HzRG850.1 (Table~1) suggest that it is
resolved. Aperture-corrected fluxes (based on Mars, with a 4.5$''$
diameter) increase with aperture size, showing a 50\% increase in flux
when going from a 5$''$- to a 15$''$-diameter aperture. This indicates
that HzRG850.1 subtends $>4.5''$, consistent with the result of
deconvolving it from the best-fit Gaussian profile for the
instrumental beam which suggests a FWHM of $\sim$10$''$. The
850-$\mu$m emission from 4C\,41.17 and HzRG850.2 may also be resolved;
their beam-corrected FWHM are $\sim 7''$ and $\sim 2''$ respectively.

Two other 850-$\mu$m sources are also {\it marginally} detected
($\ge3\sigma$) in the cleaned map (one of these is also seen in the
uncleaned map) and one further source is detected towards the eastern
edge of the larger, uncleaned map: it reaches 6$\sigma$ at its peak
and it is seen in the E--W- and N--S-chopped sub-images so we consider
it a robust detection, though its proximity to the map edge means that
caution should be exercised regarding its precise flux and position
(again, see Table~1).

Of the four robust 850-$\mu$m sources, two are also detected at
450$\,\mu$m: HzRG850.1 and 4C\,41.17. The positional coincidence with
the 850-$\mu$m sources leaves little doubt of their reality, whilst
the low flux levels indicate high redshifts. HzRG850.2 is not
detected, although it lies on the extreme edge of the 450-$\mu$m
map. Neither of the two marginal 850-$\mu$m sources are detected at
450$\,\mu$m, and the bright eastern source does not lie on the
450-$\mu$m map. One further faint ($3\sigma$) source appears in the
450-$\mu$m map; it has no counterpart at 850\,$\mu$m.

\subsection{Source counts and clustering}

Clearly, this field shows startling evidence for extremely luminous
submm galaxies, possibly in a structure associated with the radio
galaxy, 4C\,41.17 at $z=3.8$.  We first discuss the strength of this
over-density and the evidence for excesses of other classes of
proposed high-redshift galaxies in this field, before going on to
examine the available information on the redshifts of the submm
galaxies to determine if it is consistent with them lying at the same
redshift as 4C\,41.17 and hence whether they can be placed in the
heart of the proposed proto-cluster.

How does the submm galaxy density towards 4C\,41.17 compare with that
expected in blank-fields (Smail et al.\ 1997; Hughes et al.\ 1998;
Barger et al.\ 1999; Blain et al.\ 1999; Eales et al.\ 1999)?  {\it
Excluding} 4C\,41.17, the density of $S(850\,\mu{\rm m}) \ge 8\,$-mJy
sources is $1220\pm 860$\,deg$^{-2}$; the weighted mean 8-mJy
blank-field count is an order of magnitude lower at $134\pm
57$\,deg$^{-2}$. Delving slightly deeper, $S(850\,\mu{\rm m}) \ge
6.5\,$-mJy is $1830\pm 860$\,deg$^{-2}$ versus $303\pm108$\,deg$^{-2}$
in blank fields. The observed source density at $S(850\,\mu{\rm m})
\ge 8$-mJy in the 4C\,41.17 is equal to that seen in blank fields at a
flux limit of only $S(850\,\mu{\rm m}) \sim 2.5\,$mJy, below our
4$\sigma$ detection threshold.\footnote{This allows us to rule out the
alternative suggestion that the high density of submm galaxies results
from amplification bias due to a foreground mass structure, which also
amplifies the radio source. To achieve the necessary amplification
across the SCUBA field ($A\sim 3$), the lens would have to have a mass
of $\sim 10^{15}M_\odot$ within the field -- such a massive foreground
cluster would be easily visible.}

Adopting the standard probabilistic methodology used in studies of
clustering and confusion (e.g.\ Lilly et al.\ 1999), it is a simple
exercise to determine the likelihood, $P$, of detecting so many bright
sources within the 4C\,41.17 field: $P<0.00166$, meaning that on
average we would have to observe more than 600 typical blank fields
before finding a configuration of the type seen towards
4C\,41.17. Note that $P$ is given as an upper limit because HzRG850.1
is the brightest known 850-$\mu$m source and $S(850\,\mu{\rm m}) \ge
15\,$mJy is thus unknown; we conservatively adopt [$S(850\,\mu{\rm m})
\ge 15\,$mJy] = [$S(850\,\mu{\rm m}) \ge 8\,$mJy].

We note that the integrated 850-$\mu$m flux density in sources
brighter than 2\,mJy in this field is $\sim 3.3\times
10^{-10}$\,W\,m$^{-2}$\,sr$^{-1}$, so we have resolved $\gs 90$\% of
the expected extragalactic background in the submm, based on its
detection by {\it COBE} (Fixsen et al.\ 1998), and would far exceed it
by $\sim 1$\,mJy.

\subsection{Counterparts}

Before discussing the optical and near-IR counterparts to the submm
sources around 4C\,41.17, we introduce a nomenclature for their
classification, analogous to that used for proto-stars, and building
upon the evolutionary scheme for ULIRGs proposed by Sanders et al.\
(1988).  For operational purposes we base this scheme on the typical
depths achieved in follow-up observations ($I\sim 26$, $K\sim 21$) and
the properties of the submm galaxies discussed in Ivison et al.\
(1998, 2000), Hughes et al.\ (1998), Smail et al.\ (1999) and Soucail
et al.\ (1999). We define the following classes: Class~0, very-highly
obscured sources, where there is no plausible optical or near-IR
counterpart; Class~I, highly obscured sources, where only a near-IR
counterpart exists (often EROs --- Elston, Rieke \& Rieke 1988; Dey et
al.\ 1999; Smail et al.\ 1999); and Class~II, where an obvious optical
counterpart is seen (IIa: pure starburst; IIb: type-II AGN; IIc:
type-I AGN).  The latter class may overlap with the most massive
examples of Lyman-break-selected objects (Adelberger \& Steidel 2000).
We note that Classes 0 and I may also include sources which are
fainter in the optical/near-IR by virtue of being at higher redshifts;
however, for the purposes of the discussion here we assume that the
different classifications arise solely from differences in the
obscuration of the sources.

Turning to the optical and near-IR imaging of this field, we start by
discussing HzRG850.1.  There are three possible optical or near-IR
counterparts to this source lying within or near the 450- and
850-$\mu$m errors circles (Figs.~2 \& 3; see also Table~2).  The
bluest of these sources is 850.1.K1 (Class~II), which falls within the
450-$\mu$m error circle. Lacy \& Rawlings (1996) detect this source in
the $U$-band ($U-V \sim -0.9$; LR3 in their paper) and we therefore
conclude that 850.1.K1 is probably a low-redshift, blue galaxy and
unlikely to be associated with the SCUBA source. The remaining two
galaxies, 850.1.K2 and 850.1.K3 (both Class~I) have similar very red
colors, $V-K \sim 7$, with 850.1.K2 falling within both the 850-$\mu$m
and 450-$\mu$m error circles.  The similar, very red colors of
850.1.K2 and 850.1.K3 (Fig.~3; Table~2), combined with the fact that
the 850-$\mu$m source HzRG850.1 is spatially extended, suggests that
both these galaxies may be jointly responsible for the submm
emission. We also note some {\it very} faint emission in $V$ and $K$
to the west of 850.1.K2, and in $V$ alone to the south.

For HzRG850.2, three faint galaxies are visible in the $K$ image
(Figs.~2 \& 3): the bluer of these, 850.2.K1, lies just within the
error ellipse, while the others, 850.2.K2 and 850.2.K3, are much
redder, $R-K \ge 6.4$, and lie on the edge of the nominal error
ellipse.  In the deep {\it HST} WFPC2 F702W image, both 850.2.K1 and
850.2.K2 are each resolved into two components, while 850.2.K3 is
undetected ($R>26$, $3\sigma$). For 850.2.K1 the sub-components both
appear to be galaxies, while in 850.2.K2 they may either be two
galaxies or a single system crossed by a dust lane.  As with
HzRG850.1, it is plausible that both of these counterparts may be
contributing to the submm emission from HzRG850.2, given its
morphology (Fig.~1).

%
%
{\scriptsize
\begin{center}
{\sc Table 2}

\vspace{0.1cm}
{\sc Candidate Near-IR Identifications for Submm Sources}

\vspace{0.3cm}
\begin{tabular}{lccccc}
\hline\hline
\noalign{\smallskip}
ID  &  R.A.\ (J2000) & Dec.\ (J2000) & $K$  & $V-K$ & $R-K$ \cr
\hline
\noalign{\smallskip}
850.1.K1 & 06 50 51.12 & +41 30 09.2 &  20.0$\pm$0.4 & 4.5$\pm$0.4 & ... \cr
850.1.K2 & 06 50 51.05 & +41 30 06.8 &  19.0$\pm$0.2 & 7.0$\pm$0.3 & ... \cr
850.1.K3 & 06 50 51.29 & +41 29 58.6 &  19.4$\pm$0.3 & 6.9$\pm$0.3 & ... \cr
850.2.K1 & 06 50 49.14 & +41 31 25.6 &  19.6$\pm$0.3 & 4.6$\pm$0.3 & 4.4$\pm$0.3 \cr
850.2.K2 & 06 50 49.07 & +41 31 23.7 &  19.2$\pm$0.2 & 7.4$\pm$0.4 & 6.4$\pm$0.2 \cr
850.2.K3 & 06 50 49.71 & +41 31 28.2 &  20.2$\pm$0.4 & $5\sigma>$6.1 & $5\sigma>5.3$ \cr
850.3    &     ...     &     ...     & $5\sigma>20.0$   &   ...    & ... \cr
850.4.K1 & 06 50 53.95 & +41 29 40.4 &  20.5$\pm$0.5    &   ...    & $5\sigma>$5.6 \cr
850.5    &     ...     &     ...     & $5\sigma>20.0$   &   ...    & ... \cr
\noalign{\smallskip}
\noalign{\hrule}
\noalign{\smallskip}
\end{tabular}
\end{center}
}

HzRG850.3 unfortunately lies in a region where the $K$-band imaging is
relatively shallow (Fig.~2). No obvious counterpart is visible in the
KPNO $K$-band image down to $K=19.8$ ($3\sigma$).  There is, however,
a very faint $V\gs 27$ source visible within the 850-$\mu$m error
circle. In addition, there are a pair of faint ($V\sim 25.6$, 26.0)
galaxies that lie $\sim 5.6\arcsec$ northeast of the SCUBA source
centroid. However, the faintness of these galaxies and the lack of any
useful color information renders it difficult to place much confidence
in these galaxies as likely candidates for the submm emission.  We
conclude that HzRG850.3 must either be a Class~0 or I submm galaxy.

The remaining SCUBA sources lie in comparatively shallow regions of
the near-IR and optical images and do not show unambiguous
identifications.  HzRG850.4 has a very faint ($K\sim 20.5$) and
extended counterpart (850.4.K1) visible in the $K$-band image within
the 850-$\mu$m error circle (Fig.~2).  The position of 850.4.K1 in the
$V$-band image is contaminated by the diffraction spike from a bright
star, but there is no obvious counterpart to this source visible in
the deep {\it HST} F702W image and we conclude that it is a Class~I
source. For HzRG850.5 there are no obvious sources visible in the $K$
or F702W image (Class 0 or I), and the $V$-band image is again
contaminated by the nearby star.

In summary, HzRG850.1 and HzRG850.2 are likely to be associated with
EROs (i.e.\ Class I sources). HzRG850.3 and HzRG850.4 may be
associated with very faint galaxies, and HzRG850.5 has no obvious
counterpart (Class 0 or I).  It is interesting that our success in
obtaining candidate identifications, in particular the association
with EROs, correlates with the observed flux density of the submm
emission. Is it possible that the brightest submm sources are
typically associated with Class~I or II counterparts (e.g.\ HR10, Dey
et al.\ 1999; SMM\,J09429+4658, Smail et al.\ 1999;
SMM\,J02399$-$0136, Ivison et al.\ 1998; SMM\,J14011+0252, Ivison et
al.\ 2000) whereas fainter submm sources typically have Class~0
counterparts (see also Hughes et al.\ 1998; Smail et al.\ 2000).

\subsection{Redshift constraints}

Finally, we discuss whether the submm sources can be unambiguously
placed at high redshifts, $z\gg 1$, and hence whether we can
strengthen the case for their association with 4C\,41.17. In the
absence of deep spectroscopy of the counterparts discussed in \S3.3,
we are reliant on other redshift-sensitive parameters. Two examples
are: $a$) the 450- to 850-$\mu$m flux ratio, which falls steadily as
the peak of the dust SED shifts through the 450-$\mu$m filter at
increasing redshift; and $b$) the 850-$\mu$m to 1.4-GHz flux ratio,
based upon the local far-IR/radio correlation for star-forming
galaxies, is a robust redshift indicator where $F_{\nu} \propto
\nu^{+3.5}$ at submm wavelengths, while $F_{\nu} \propto \nu^{-0.7}$
in the radio, yielding a flux ratio that rises initially as
$(1+z)^{+4.2}$ (Carilli \& Yun 1999, 2000; Blain 1999; see also Smail
et al.\ 2000).

Dealing with the 450- to 850-$\mu$m flux ratio first, the observed
ratios for 4C\,41.17 and HzRG850.1 are $3.2\pm 0.8$ and $2.2\pm 0.6$,
consistent with their being at the same redshift, with best-fit values
and ranges of $z \sim 3.8$ ($2.8 < z < 5.1$) and $\sim 4.8$ ($3.5 < z
< 6.1$), respectively, based on extreme models of starbursts (Hughes
et al.\ 1998). These limits are supported by constraints from the
850-$\mu$m to 1.4-GHz flux ratio. For HzRG850.1, where $S_{850\mu{\rm
m}}/S_{1.4{\rm GHz}} \ge 180$ (taking radio limits of $3\sigma <
90$\,$\mu$Jy from the re-reduced radio image published originally by
Carilli, Owen \& Harris 1994), we arrive at a robust limit of $z\ge
2$. The limits for HzRG850.2 and HzRG850.3 are slightly weaker, but
they are certainly not local galaxies.

The extreme colors, $V-K \sim 7$, of the probable galaxy counterparts
to HzRG850.1 and HzRG850.2, as well as the faintness of the
counterparts to the remaining sources, suggests that these systems
also lie at high redshifts.  The $K$-band magnitudes for the candidate
identifications quoted are fairly bright: if these galaxies do lie at
$z=3.8$ then they are very luminous, in excess of 50$L^*$, although we
note that luminosities close to this are found for the confirmed Class
II counterparts SMM\,J02399$-$0136 ($z=2.8$, Ivison et al.\ 1998) and
SMM\,J14011+0252 ($z=2.6$, Ivison et al.\ 2000).

We conclude that the available constraints suggest that the
over-density of bright submm galaxies identified in the 4C\,41.17
field is likely to lie at $z>2.8$ and hence is consistent with being
associated with the radio galaxy at $z=3.8$ (though spectroscopic
confirmation is clearly a top priority).  This suggests that in
addition to the proposed over-density of Lyman-break galaxies
identified around 4C\,41.17 by Lacy \& Rawlings (1996), we should also
add a comparably numerous population of highly-obscured and very
luminous submm galaxies.

\section{Conclusions}

We report the discovery of five new luminous submm sources within a
$2.5'$-diameter (1\,Mpc at $z=3.8$) area centered on the submm-bright
$z=3.8$ radio galaxy, 4C\,41.17. Three of these sources are as bright
or brighter than any other known blank-field submm galaxies.  This
surface density of submm galaxies is an order of magnitude greater
than that expected from blank-field counts: we would need to map at
least 600 blank fields to find a chance configuration of such bright
sources. Using the available constraints, we suggest that this
over-density lies at $z>2.8$ and is therefore consistent with a
structure associated with 4C\,41.17.

We note that this field exhibits an over-density above that expected
in blank fields, not only at submm wavelengths, but also in samples of
galaxies selected via the Lyman-break technique (Lacy \& Rawlings
1996) and for their extremely red colors (EROs).  As many as five of
the latter class of galaxies may be directly associated with the
luminous submm sources seen in this field.

We introduce a classification scheme for the counterparts of submm
sources, based upon observations of well-studied SCUBA galaxies.
Within the framework of this scheme we propose that submm galaxies and
ERO starbursts represent different aspects of a single evolutionary
cycle.  Following on from this, we suggest that the over-densities of
both the submm and ERO populations in this field represent young,
dusty starbursts forming within a proto-cluster centered on the radio
galaxy at $z=3.8$, which also hosts a population of less-obscured
Lyman-break galaxies.

More observations are clearly needed to confirm the nature of this
over-density, to test our suggestion that it is associated with the
radio source at $z=3.8$ and, most interestingly, to investigate if it
represents a virialised structure.  The most feasible test of the
latter issue will be to image this field using the {\it Chandra} or
{\it Newton} X-ray observatories to identify emission from the hot
intracluster medium confined within the potential well of the proposed
proto-cluster.

We are currently expanding our survey to cover 15 fields centered on
HzRGs to determine more reliable limits on the prevalence of similar
over-densities of submm sources around massive galaxies at high
redshifts, $z=3$--5.

\section*{Acknowledgements}

We thank Chris Carilli and Wil van Breugel for providing their Very
Large Array and {\it HST} images of the 4C\,41.17 field, Dr F.\
Chaffee for assistance with obtaining the $V$-band image, Gary Punawai
and Greg Wirth for their expert assistance at the W.\,M.\ Keck
Observatory, as well as H.\ Halbedel, G.\ MacDougal, C.\ Snedden, H.\
Schweiker and G.\ Tiede for their expert assistance at the KPNO
2.1m. We are grateful to the referee, Chris Carilli, for suggestions
that improved this paper markedly. The JCMT is operated by the Joint
Astronomy Centre on behalf of the United Kingdom Particle Physics and
Astronomy Research Council (PPARC), the Netherlands Organisation for
Scientific Research, and the National Research Council of Canada.
KPNO is a division of the National Optical Astronomy Observatory,
which is operated by the AURA, under cooperative agreement with the
NSF.  The W.\,M.\ Keck Observatory is operated as a scientific
partnership among the California Institute of Technology, the
University of California and NASA.

\end{document}